# Electron tomography at 2.4 Å resolution

M. C. Scott[1*], Chien-Chun Chen[1*], Matthew Mecklenburg[1*], Chun Zhu[1], Rui Xu[1], Peter Ercius[2], Ulrich Dahmen[2], B. C. Regan[1] & Jianwei Miao[1]

[1]*Department of Physics and Astronomy and California NanoSystems Institute, University of California, Los Angeles, CA 90095, USA.* [2]*National Center for Electron Microscopy, Lawrence Berkeley National Laboratory, Berkeley, CA 94720, USA.*
*These authors contributed equally to this work.*
*Correspondence and requests for materials should be addressed to J. M. (miao@physics.ucla.edu).*

**Transmission electron microscopy (TEM) is a powerful imaging tool that has found broad application in materials science, nanoscience and biology[1-3]. With the introduction of aberration-corrected electron lenses, both the spatial resolution and image quality in TEM have been significantly improved[4,5] and resolution below 0.5 Å has been demonstrated[6]. To reveal the 3D structure of thin samples, electron tomography is the method of choice[7-11], with resolutions of ~1 nm³ currently achievable[10,11]. Recently, discrete tomography has been used to generate a 3D atomic reconstruction of a silver nanoparticle 2-3 nm in diameter[12], but this statistical method assumes prior knowledge of the particle's lattice structure and requires that the atoms fit rigidly on that lattice. Here we report the experimental demonstration of a general electron tomography method that achieves atomic scale resolution without initial assumptions about the sample structure. By combining a novel projection alignment and tomographic reconstruction method with scanning transmission electron microscopy, we have determined the 3D structure of a ~10 nm gold nanoparticle at 2.4 Å resolution. While we cannot definitively locate all of the atoms inside the nanoparticle, individual atoms are observed in some regions of the particle and several grains are identified at three dimensions. The 3D surface morphology and internal lattice structure revealed are consistent with a distorted icosahedral multiply-twinned particle. We anticipate that this general method can be applied not only to determine the 3D structure of nanomaterials at atomic scale resolution[13-15], but also to improve the spatial resolution and image quality in other tomography fields[7,9,16-20].**

Since its introduction in 1968, electron tomography has been primarily used to determine the 3D structure of biological samples[7,9]. In the last decade, electron tomography has been increasingly applied in materials science and nanoscience through the use of scanning transmission electron microscopy (STEM)[8,10,11] The highest resolution presently achieved by STEM tomography is around 1 nm in three dimensions[10,11], although slightly higher resolution has been obtained in a study of fullerene-like nanostructures with bright-field electron tomography[21]. A general electron tomography method with atomic scale resolution, however, has not been demonstrated for several reasons. First, aligning the projections of a tomographic tilt series to a common axis with atomic level precision is technically challenging. Second, radiation damage limits the number of projections that can be acquired from a single object[13,22]. Finally, specimens cannot usually be tilted beyond ±79°, preventing acquisition of data from the "missing wedge"[7-11]. Here we demonstrate that these limitations can be overcome or alleviated by applying a novel alignment approach and an iterative tomographic reconstruction method to a tilt series obtained via annular dark field

(ADF)-STEM[23,24].

The conventional alignment approach used in electron tomography either relies on fiducial markers such as colloidal gold beads or is based on the cross-correlation between neighboring projections[7,9]. To our knowledge, neither of these alignment approaches can achieve atomic level precision. To overcome this limitation, we have developed a method based on the center of mass (CM), which is able to align the projections of a tilt series at atomic level accuracy even with noise (Methods). To address the other two difficulties, we have implemented a data acquisition and tomographic reconstruction method, termed equally sloped tomography (EST)[25,16,18-20]. Compared to conventional tomography that reconstructs a 3D object from a tilt series of projections with constant angular increments, EST acquires a tilt series with equal slope increments, and then iterates back and forth between real and reciprocal space (Methods). In each iteration, constraints such as the sample boundary (*i.e.* support) and positivity of the Coulomb potential are applied in real space, while the measured projectional slices (*i.e.* the Fourier transform of the projections) are enforced in reciprocal space. Each iteration is monitored by an error metric, and the algorithm is terminated after reaching a maximum number of iterations.

To test the feasibility of achieving a high resolution tomographic reconstruction by the CM and EST methods, we conducted numerical simulations on a ~5 nm gold nanoparticle with icosahedral symmetry and a total of 3871 atoms (Supplementary Figs. 3a, 4a, c and e). A tilt series of 55 projections was obtained from the particle using multislice STEM calculations[26] (energy: 300 keV, spherical aberration: 1.2 mm, illumination semi-angle: 7.98 mrad, defocus: 48.6 nm, detector inner and outer angles: 13 and 78 mrad). To minimize non-linear intensity contributions caused by dynamical scattering and electron channeling[27], projections along zone axis orientations were avoided. The tilt angles range from -72.6° to +72.6° with equal slope increments. To more closely approximate realistic experimental conditions, several additional modifications were made to generate the simulated data. First, the tilt angles were continuously shifted from 0° to 0.5° over the process of the tilt series and the magnification of the images was continuously changed from 0 to 0.2%. Second, each projection in the tilt series was arbitrarily shifted along the X- and Y-axes, where the electron beam direction is parallel to the Z-axis. Finally, Poisson noise was added to the projections of the tilt series with a total electron dose of $6.1\times10^6$ e/Å$^2$. Supplementary Fig. 3 shows a linear projection of the model at 0° and the corresponding multislice STEM projection. The apparent increase of the atom size in the multislice projection was mainly caused by the non-linear and diffraction effects in the nanoparticle.

The 55 projections were aligned to a common tilt axis with the CM method, and were then reconstructed with the EST algorithm. Supplementary Figs. 4b, d, and f show three 2.5 Å thick central slices of the 3D reconstruction in the XY, ZX and ZY planes. Although the missing wedge problem is not completely solved (the top and bottom parts in Supplementary Fig. 4f) and the size of the reconstructed atoms is increased mainly due to the non-linear and diffraction effects, the atomic positions and grain boundaries in the 3D reconstruction are consistent with the model. The simulation results indicate that the CM and EST methods can be used to achieve an atomic scale resolution reconstruction from a tilt series of 55 projections with a missing wedge, non-linear effects, Poisson noise and experimental errors.

Next, the CM and EST methods were applied to experimental tilt series acquired from gold nanoparticles. Nanoparticles are an important class of materials with properties different from either molecules or bulk solids[13-15], and nano-gold is among the most widely studied of this class of material due to its broad applications in chemistry, biology, materials science, nanoscience and nanotechnology[28]. In this study, we imaged gold particles with a diameter of ~10 nm as smaller particles are not as stable under an electron beam[13,22]. To reduce the electron dose, we used a low exposure data acquisition scheme with a 300 keV ADF-STEM (Methods). When focusing an image, a nearby nanoparticle was first viewed (not the particle of interest), thus reducing the unnecessary radiation dose to the particle under study. Using this scheme, we acquired several tomographic tilt series of gold nanoparticles. Supplementary Fig. 5 shows a tilt series of 69 projections and their Fourier transforms, with a total electron dose of ~7.6×10^6 e/Å^2. Supplementary Fig. 6 shows three 0° projections and their Fourier transforms measured during the acquisition of this tilt series to monitor the effects of radiation damage. While some minor shape changes occurred, the crystal lattice structure of the particle remained consistent throughout the experiment. To investigate the non-linear effects in the experiment, we simulated a ~10 nm gold particle with icosahedral symmetry and performed multislice STEM calculations on an 11.5 Å thick slab of this particle[26] (Supplementary Fig. 7). Although the atom size was increased mainly due to the non-linear diffraction and dynamical scattering effects, the multislice STEM projection exhibits consistent crystal lattice structure with the model. We then calculated a tilt series for a 2.5 Å central slice using the multislice simulations. The tilt series consists of 69 projections with a tilt range of ±72.6°. Supplementary Fig. 8 shows the model and reconstructed slices. The atomic positions and the internal grain boundaries are resolved except in very few places due to the non-linear effects in the projections.

After investigating the non-linear effects, we performed post data analysis of the experimental tilt series (Methods), and aligned the projections with the CM method. To reconstruct the 3D structure, we first estimated a loose 3D support, defined to be larger than the particle boundary. After 500 iterations of the EST algorithm, the reconstruction was used to determine a tight support (*i.e.*, close to the true boundary of the particle). Using the tight support, we ran another 500 iterations and obtained a final 3D structure. To examine the quality of the reconstruction, we calculated 69 projections from the final 3D structure and found the average normalized discrepancy with the measured projections to be 6.7% (Supplementary Methods and Supplementary Tab. 1). Representative measured and calculated projections at different particle orientations are shown in Fig. 1 and Supplementary Fig. 9. While there are some minor differences between the two projections, the overall shape and lattice structure agree well. To more rigorously examine the accuracy of the reconstruction, an EST reconstruction was performed from 68 experimental projections by removing the 7.1° projection. The 3D reconstruction was then projected back to calculate the projection at 7.1°, which is consistent with the experimentally measured one (Supplementary Fig. 10).

To estimate the resolution achieved in the reconstruction, we chose a 3.36 Å thick central slice in the XY plane. Figs. 2a and b show the slice and its Fourier transform in which the distance between two neighboring atom columns and the location of the Bragg peaks indicate that a resolution of 2.4 Å was achieved in the X and Y directions. To estimate the resolution along the Z-axis (beam direction), we selected a 3.36 Å thick slice with the horizontal axis along the Z-axis (Figs. 2c and d). The resolution close to the Z-axis was determined to be 2.4 Å. Individual atoms are visible in some regions of the slices, but not all atoms can be identified in the slices.

Supplementary Fig. 11 shows two 3.36 Å thick slices of the nanoparticle in a different orientation, exhibiting crystal lattice structure that is not present in Fig. 2a and c. The apparent flattening of the particle along the beam axis was also observed in the 3D reconstructions (Fig. 2 and Supplementary Movie 1), and was likely caused by the interaction between the nanoparticle and the Si substrate.

To visualize the internal structure and the morphology of the gold nanoparticle, we generated 3D volume and iso-surface renderings of the reconstruction, in which both surface and internal lattice structures are visible (Supplementary Movies 1 and 2). Fig. 3a and b show volume renderings of the nanoparticle and their Fourier transforms (insets) at the 2- and 3-fold symmetry orientations. The corresponding iso-surface renderings at the same orientations are shown in Figs. 3c and d. The overall 3D shape and facets of the nanoparticle are consistent with an icosahedron (insets in Figs. 3c and d). To identify internal 3D grains, we applied the 3D Fourier transform to the reconstruction. By identifying the Bragg peaks of each major grain and applying the 3D inverse Fourier transform to the selected Bragg peaks, we determined four major 3D grains inside the gold nanoparticle (Methods). Fig. 4 and Supplementary movie 3 show a volume rendering of the four grains in three dimensions, in which grains 1, 2 and grains 3, 4 are related by mirror-reflection across the horizontal interfaces marked by dotted lines. The angle enclosed by close-packed planes across these interfaces was measured to be 69.9°±0.8° between grains 1 and 2, and 71.3°±0.8° between grains 3 and 4, both of which are consistent with the angle for an fcc twin boundary (70.53°). By applying the same method to some other Bragg peaks, we identified 3D grains in the top and bottom parts of the particle (Supplementary Fig. 12). The surface morphology and the internal lattice structures suggest that this is a distorted icosahedral multiply-twinned particle, typically found for nano-gold in the size range above 10 nm[29].

By combining the CM alignment technique and the EST reconstruction method with an ADF-STEM, we have determined the 3D structure of a ~10 nm gold nanoparticle at 2.4 Å resolution from a tilt series of 69 projections with a missing wedge. Several grains are identified inside the nanoparticle in three dimensions. While individual atoms are visible in some regions of the nanoparticle, we cannot definitively locate all of the atoms inside the particle. In order to identify all the atoms in the particle (estimated to be ~23800) without using atomicity and bond information, a resolution higher than 2.4 Å is needed, which requires future developments. With aberration-corrected STEM[4,23,24], better 3D resolution and image quality should be achievable, but extended depth-of-field techniques may have to be applied to the tilt series before the EST reconstruction can be performed. Compared to atom-probe tomography[30], this non-destructive technique can not only handle isolated nanoparticles, but also provide the 3D local structure of complex nanomaterials at high resolution.

## METHODS SUMMARY

Gold nanoparticles with size of ~10 nm (Ted Pella) were supported on 5 nm thick Si membranes (TEMwindows.com) which can withstand plasma cleaning for a longer period than carbon substrates to alleviate carbon contamination. Tomographic tilt series with equal slope increments were acquired from the gold nanoparticles by using an ADF-STEM (FEI Titan 80-300). The tilt angles ($\theta$) were determined by[25,16] $\theta = -\tan^{-1}[(N + 2 - 2n)/N]$ for $n = 1, \ldots, N$ and $\theta = \pi/2 - \tan^{-1}[(3N + 2 - 2n)/N]$ for $n = N+1, \ldots, 2N$ with $N = 32$ or 64 in this experiment. The angles beyond ±72.6° were not accessible due to the geometry of the support grid. To monitor the radiation induced stability effect, several projections at the same particle orientation were measured during the acquisition of each tilt series (Supplementary Fig. 6). To improve the 3D reconstruction and enhance the signal to noise ratio, the background of the projections was subtracted and 2×2 pixel binning was performed for each projection. After post data analysis, the tilt series was aligned with the CM method (Methods). The reconstruction of the aligned tilt series was conducted using the EST method, which iterated back and forth between real and reciprocal space with constraints enforced in real space and measured data in reciprocal space (Methods, Supplementary Methods and Supplementary Fig. 2). To examine the reconstruction quality, the reconstructed 3D structure was projected back to obtain 69 projections, which were compared to the corresponding measured ones. An average $R_{real}$ (Supplementary Methods) was calculated to be 6.7%, indicating a good quality reconstruction (Fig. 1, Supplementary Tab. 1).

**References**

1. Williams, D. B. & Carter, C. B. *Transmission Electron Microscopy: A Textbook for Materials Science* 2nd ed. (Springer, 2009).

2. Spence, J. C. H. *Experimental High-Resolution Electron Microscopy* 3rd ed. (Oxford University Press, New York, 2003).

3. Frank, J. *Three-Dimensional Electron Microscopy of Macromolecular Assemblies.* (Oxford University Press, USA, 2006).

4. Batson, P. E., Dellby, N. & Krivanek, O. L. Sub-ångstrom resolution using aberration corrected electron optics. *Nature* **418**, 617-620 (2002).

5. Haider, M., *et al.* Electron microscopy image enhanced. *Nature* **392**, 768-769 (1998).

6. Erni, R., Rossell, M. D., Kisielowski, C. & Dahmen, U. Atomic-resolution imaging with a sub-50-pm electron probe. *Phys. Rev. Lett.* **102**, 096101(2009).

7. Frank, J. *Electron Tomography* (Plenum, New York, 1992).

8. Midgley, P. A. & Weyland, M. 3D electron microscopy in the physical sciences: the development of Z-contrast and EFTEM tomography. *Ultramicroscopy* **96**, 413–431 (2003).

9. Lučić, V., Förster, F. & Baumeister, W. Structural studies by electron tomography: from cells to molecules. *Annu. Rev. Biochem.* **74**, 833-865 (2005).

10. Midgley, P. A. & Dunin-Borkowski, R. E. Electron tomography and holography in materials science. *Nature Materials* **8**, 271-280 (2009).

11. Arslan, I., Yates, T. J. V., Browning, N. D. & Midgley, P. A. Embedded nanostructures revealed in three dimensions. *Science* **309**, 2195-2198 (2005).

12. Van Aert, S., Batenburg, K. J., Rossell, M. D., Erni, R. & Van Tendeloo, G. Three-dimensional atomic imaging of crystalline nanoparticles. *Nature* **470**, 374–377 (2011).

13. Marks, L. D. Experimental studies of small particle structures. *Rep. Prog. Phys.* **57**, 603-649 (1994).

14. Billinge, S. J. L. & Levin, I. The problem with determining atomic structure at the nanoscale. *Science* **316**, 561-565 (2007).

15. Yacaman, M. J., Ascencio, J. A., Liu, H. B. & Gardea-Torresdey, J. Structure shape and stability of nanometric sized particles. *J. Vac. Sci. Technol. B* **19**, 1071-1023 (2001).

16. Lee, E. *et al.* Radiation dose reduction and image enhancement in biological imaging through equally sloped tomography. *J. Struct. Biol.* **164**, 221–227 (2008).

17. Kak, A.C. & Slaney, M. *Principles of Computerized Tomographic Imaging* (SIAM, Philadelphia, 2001).

18. Fahimian, B. P., Mao, Y., Cloetens, P., & Miao, J. Low dose X-ray phase-contrast and absorption CT using equally-sloped tomograph. *Phys. Med. Bio.* **55**, 5383-5400 (2010).

19. Mao, Y., Fahimian, B. P., Osher, S. J., & Miao, J. Development and optimization of regularized tomographic reconstruction algorithms utilizing equally-sloped tomography. *IEEE Trans. Image Processing* **19**, 1259-1268 (2010).

20. Jiang, H. *et al.* Quantitative 3D imaging of whole, unstained cells by using X-ray diffraction microscopy. *Proc. Natl. Acad. Sci. USA* **107**, 11234–11239 (2010).

21. Bar Sadan, M. *et al.* Toward atomic-scale bright-field electron tomography for the study of fullerene-like nanostructures. *Nano Lett.* **8**, 891-896 (2008).

22. Bovin, J. -O., Wallenberg, R. & Smith, D. J. Imaging of atomic clouds outside the surfaces of gold crystals by electron microscopy. *Nature* **317**, 47-49 (1985).

23. D. A. Muller. Structure and bonding at the atomic scale by scanning transmission electron microscopy. *Nature Materials* **8**, 263-270 (2009).

24. Pennycook, S. J. & Nellist, P. D. *Scanning Transmission Electron Microscopy: Imaging and Analysis* 1st ed. (Springer, 2011).

25. Miao, J., Föster, F. & Levi, O. Equally sloped tomography with oversampling reconstruction. *Phys. Rev. B* **72**, 052103 (2005).

26. Kirkland, E. J. *Advanced Computing in Electron Microscopy* 2nd ed. (Springer, 2010).

27. Howie, A. Diffraction channelling of fast electrons and positrons in crystals. *Phil. Mag.* **14**, 223-237 (1966).

28. Daniel, M. C. & Astruc, D. Gold nanoparticles: assembly, supramolecular chemistry, quantum-size-related properties, and applications toward biology, catalysis, and nanotechnology. *Chem. Rev.* **104**, 293-346 (2004).

29. Barnard, A. S., Young, N. P., Kirkland, A. I., van Huis, M. A. & Xu, H. Nanogold: a quantitative phase map. *ACS Nano* **3**, 1431–1436 (2009).

30. Arslan, I., Marquis, E.A., Homer, M., Hekmaty, M. & Bartelt, N.C. Towards better 3-D reconstructions by combining electron tomography and atom-probe tomography. *Ultramicroscopy* **108**, 1579-1585 (2008).

**Acknowledgements** We thank E. J. Kirkland for help with multislice STEM calculations, R. F. Egerton, Z. H. Zhou and J. A. Rodríguez for stimulating discussions and I. Atanasov for assistance in data acquisition. The tomographic tilt series were acquired at the Electron Imaging Center for NanoMachines of California NanoSystems Institute. This work was partially supported by UC Discovery/TomoSoft Technologies (IT107-10166). J.M. thanks partial support by the National Institutes of Health (GM081409-01A1).

## Figure legends

**Figure 1**. Representative measured (**a**) and calculated (**c**) projections and their Fourier transforms (**b,d**) at 7.1°, where insets show the projected atomic positions inside the blue square. The overall shape of the nanoparticle and the location of the Bragg peaks agree well, indicating a good quality 3D reconstruction.

**Figure 2**. 3D resolution estimation of the reconstruction. **a** and **b**, A 3.36 Å thick central slice in the XY plane and its Fourier transform, indicating 2.4 Å resolution was achieved along the X- and Y-axes. **c** and **d**, A 3.36 Å thick slice in the ZY plane and its Fourier transform where the horizontal axis is along the Z-axis (beam direction). The resolution in the Z-axis was estimated to be 2.4 Å. Individual atoms are visible in some regions of the slices, but not all atoms can be identified in the slices.

**Figure 3**. 3D structure of the reconstructed gold nanoparticle. **a** and **b**, 3D volume renderings of the nanoparticle and their Fourier transforms (insets) at the 2- and 3-fold symmetry orientations. **c** and **d**, Iso-surface renderings of the nanoparticle at the 2- and 3-fold symmetry orientations, compared to a model icosahedron at the same orientation (insets).

**Figure 4**. Identification of four major grains inside the gold nanoparticle in three dimensions. Grains 1, 2 and grains 3, 4 are related by mirror-reflection across the horizontal interfaces marked by dotted lines. The angle enclosed by close-packed planes across these interfaces was measured to be 69.9°±0.8° between grains 1 and 2, and 71.3°±0.8° between grains 3 and 4, both of which are consistent with the angle for an fcc twin boundary (70.53°).

## METHODS

**Sample preparation.** Gold nanoparticle solution with an average particle size of ~10 nm (Ted Pella) was sonicated for ~10 minutes to prevent aggregation. The solution was then dropcast onto 5 nm thick Si membranes (TEMwindows.com). The thin membrane with size of 100×1500 μm is supported on a 100 μm thick Si frame, allowing for a tilt range of ±83°. To avoid breaking the membrane, a micromanipulator was used to place a small drop of solution onto the outer frame of the Si grid. After gently moving the drop onto the membrane, it was removed and not allowed to dry and leave excessive gold particles and contaminants. The Si grids were cleaned pre-deposition in a Gatan Solarus plasma cleaner (Model 950) for 20s using a standard $H_2/O_2$ recipe. To further ensure removal of contaminant sources, the sample holder (Fischione Model 2020) was plasma cleaned for an hour prior to data acquisition using the same recipe.

**ADF-STEM**. STEM images of gold nanoparticles were acquired on a FEI Titan 80-300 (energy: 300 keV, spherical aberration: 1.2 mm, illumination semi-angle: 7.98 mrad and defocus: 48.6 nm). The electron beam, at spot 8 with a 50 μm C2 aperture, was focused to a probe and rastered over the sample. The scattered electrons were captured by a Fischione Model 3000 ADF detector with angles between 13 mrad and 78 mrad from the optical axis. The use of ADF angles was used to improve the signal to noise ratio with a low current electron beam. The effects of non-linear image intensities and diffraction contrast were carefully determined by multi-slice simulations. The maximum tilt angles were limited by the holder to ±75°.

**Low-exposure acquisition of tomographic tilt series**. In order to reduce vibration and drift during data acquisition, the sample holder was allowed to settle for one hour after insertion into the microscope, and also for several minutes after moving to each new angle. Tilt series were acquired by manually changing the angle with equal slope increments. The tilt angles ($\theta$) were determined by[25,16] $\theta = -\tan^{-1}[(N+2-2n)/N]$ for $n = 1, \dots, N$ and $\theta = \pi/2 - \tan^{-1}[(3N+2-2n)/N]$ for $n = N+1, \dots, 2N$ with $N = 32$ or 64 in this experiment. To focus each projectional image during data acquisition, a nearby particle was used (rather than the particle of interest) to reduce the radiation dose to the particle. By using this low-exposure data acquisition scheme, we have obtained several tomographic tilt series. Supplementary Fig. 5 shows the tilt series used in this reconstruction with 69 projections and a tilt range of ±72.6°. A representative sinogram of the tilt series is shown in Supplementary Fig. 13. The probe current was ~70 pA with a dwell time of 45 μs per pixel, and the magnification of each projection was 5.2 M×. Since the pixel size in STEM mode can vary, a calibration image of the particle was taken in TEM mode, and the STEM pixel size was determined to be 0.42 Å. The total electron dose of the tilt series was estimated to be ~7.6×10$^6$ e/Å$^2$. Supplementary Fig. 6 shows three 0° projections measured during the acquisition of this tilt series. Although some minor shape changes occurred, the crystal lattices of the particle remained consistent throughout the experiment.

**Post data analysis.** In order to apply the EST method, the background surrounding the nanoparticle in each projection has to be subtracted. To systematically eliminate the background, we first projected all the projections onto the tilt axis and obtained a set of 1D curves. We then determined the optimal cut-off value for background subtraction in each projection by maximizing the cross-correlation among these 1D curves. After background subtraction, we binned 2×2 pixels into 1 pixel, which was used to enhance the signal to noise ratio in the projections and improve the EST reconstruction. The background subtracted and binned projections with pixel size of 0.84 Å were aligned using the CM approach and reconstructed with the EST method.

**The CM alignment method.** To achieve an atomic scale resolution reconstruction, the projections in a tilt series have to be aligned to a common axis (not necessarily the true tilt axis) with atomic level precision in both the X- and Y-axes where the Y-axis represents the tilt axis and Z-axis the beam direction. To align the tilt series along the Y-axis, the projections were first projected onto the Y-axis and a set of 1D curves was generated. We then chose a 1D curve at 0° as a reference, and aligned the remaining curves to the reference. To align the projections along the X-axis, we developed a method based on the center of mass (CM). When a 3D object is tilted around the Y-axis from 0° to 360°, the CM of the object forms a circle. However, in the special geometry where the CM coincides with the origin of the X-axis, this circle becomes a point. To determine the CM in this special geometry, we projected each 2D projection onto the X-axis, chose a pixel as the origin and calculated the CM along the X-axis, $x_{CM} = \sum_i x_i \rho(x_i) \Big/ \sum_i \rho(x_i)$, where $\rho(x_i)$ the Coulomb potential at position $x_i$. We then shifted this projection to set $x_{CM}$ as the new origin of the X-axis. Through repeating this process for all projections, we aligned the tilt series to the common axis that coincides with the new origin. Both our simulation and experimental results indicate that the CM alignment is a general method and can align the projections of a tilt series at atomic level accuracy, even with relatively high noise and the non-linear effects (Fig. 1, Supplementary Figs. 4, 8 and Tab. 1).

**The EST method**. When the projections of a tilt series use equal slope increments, it has been shown that a direct fast Fourier transform, the pseudopolar fast Fourier transform (PPFFT)[31], exists between a pseudopolar grid and a Cartesian grid. Supplementary Fig. 1 shows a pseudopolar grid and the PPFFT. For an $N \times N$ Cartesian grid, the corresponding pseudopolar grid is defined by a set of $2N$ lines, each line consisting of $2N$ grid points mapped out on $N$ concentric squares. The $2N$ lines are subdivided into a horizontal group (in blue) defined by $y = sx$, where $s$ is the slope and $|s| \le 1$, and a vertical group (in red) defined by $x = sy$, where $|s| \le 1$; the horizontal and vertical groups are symmetric under the interchange of $x$ and $y$, and $\Delta s = 2/N$. When these conditions are met, the PPFFT and its inverse algorithm are mathematically faithful[31]. Note that the PPFFT and its inverse algorithm were originally developed to interpolate tomographic projections from a polar to a Cartesian grid in reciprocal space. The idea of acquiring tomographic tilt-series at equal slope increments and then combining the PPFFT with iterative algorithms for 3D image reconstructions was first suggested in 2005 (ref. 25).

Compared to other data acquisition approaches such as the Saxton scheme[32], the EST data acquisition approach is different in that it acquires projections with equal slope increments in order to use the PPFFT. Although the PPFFT and its inverse provide an algebraically faithful way to do the fast Fourier transform between the Cartesian and pseudopolar grids, three difficulties limit its direct application to electron tomography. First, the tilt range has to be from -90° to +90°. Second, the number of projections in a tilt series needs to be 2N for an N × N object. Third, the grid points past the resolution circle (dashed circle in Supplementary Fig. 1) cannot be experimentally determined. We overcame these limitations by combining the PPFFT with an iterative process[25,16,18-20]. Supplementary Fig. 2 shows the schematic layout of the iterative EST method. We first convert the electron micrograph projections to Fourier slices in the pseudopolar grid. As illustrated in Supplementary Fig. 1, the distance between the sampling points on the individual 2N lines of the pseudopolar grid varies from line to line. In order to calculate the Fourier slices from the projections, the fractional Fourier transform (FrFT) is used to vary the output sampling distance of the Fourier slices[33]. By applying the inverse PPFFT, we obtain a 3D image in real space. A 3D support is defined to separate the object from a zero region where the size of the zero region is proportional to the oversampling of the projections[34]. The negative-valued voxels inside the support and the voxel values outside the support are set to zero, and a new 3D image is obtained. The forward PPFFT is applied to the new image and a set of calculated Fourier slices is obtained. We then replace the corresponding calculated Fourier slices with the measured ones, and the remaining slices are kept unchanged. The iterative process is then repeated with each iteration monitored by an $R_{recip}$ (Supplementary Methods). The algorithm is terminated after reaching a maximum number of iterations. A more detailed description of the EST method is presented in Supplementary Methods. Compared to phase retrieval in coherent X-ray/electron diffraction imaging[20,35-37], the EST method aims for solving the missing data by combining an iteration process with the PPFFT algorithm.

**Identification of the major 3D grains inside the nanoparticle**. The following procedures were used to determine the major 3D grains inside the gold nanoparticle. (i) Apply the 3D Fourier transform to the reconstructed nanoparticle and identify the Bragg peaks corresponding to a major grain. (ii) Use small spheres with soft edges to select these Bragg peaks and set other values to zero. (iii) Apply the 3D inverse Fourier transform to the selected Bragg peaks and obtain a 3D image. (iv) Convolve the 3D image with a Gaussian filter and choose a cut-off value to determine the 3D shape of the grain. (v) Use the 3D shape to identify the corresponding 3D grain in the reconstructed nanoparticle. (vi) Repeat steps (i-v) to determine other major grains.

31. Averbuch, A., Coifman, R. R., Donoho, D. L., Israeli, M. & Shkolnisky, Y. A framework for discrete integral transformations I—the pseudopolar Fourier Transform. *SIAM J. Sci. Comput.* **30**, 785–803 (2008).
32. Saxton, W. O., Baumeister, W. & Hahn, M. Three-dimensional reconstruction of imperfect two-dimensional crystals. *Ultramicroscopy* **13**, 57-70 (1984).
33. Bailey, D. H. & Swarztrauber, P. N. The fractional Fourier transform and applications. *SIAM Rev.* **33**, 389–404 (1991).
34. Miao, J., Sayre, D. & Chapman, H. N. Phase retrieval from the magnitude of the Fourier transform of non-periodic objects. *J. Opt. Soc. Am. A.* **15**, 1662-1669 (1998).
35. Miao, J., Charalambous, P, Kirz, J. & Sayre, D. Extending the methodology of X-ray crystallography to allow imaging of micrometre-sized non-crystalline specimens. *Nature* **400**, 342-344 (1999).
36. Miao, J., Ohsuna, T., Terasaki, O., Hodgson, K. O. & O'Keefe, M. A. Atomic resolution three-dimensional electron diffraction microscopy. *Phys. Rev. Lett.* **89**, 155502 (2002).
37. Zuo, J. M., Vartanyants, I., Gao, M., Zhang, R. & Nagahara, L. A. Atomic resolution imaging of a carbon nanotube from diffraction intensities. *Science* **300**, 1419-1421 (2003).

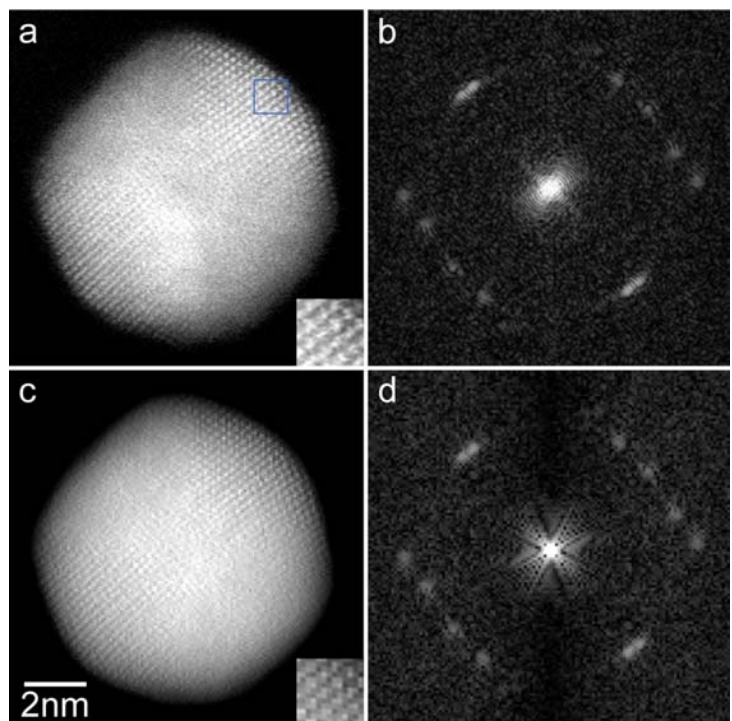

FIG. 1

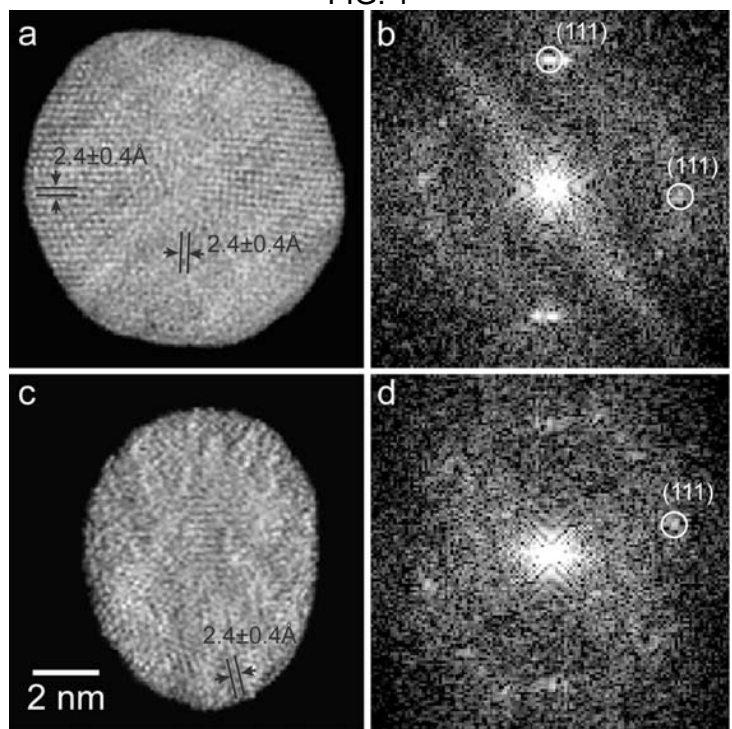

FIG. 2

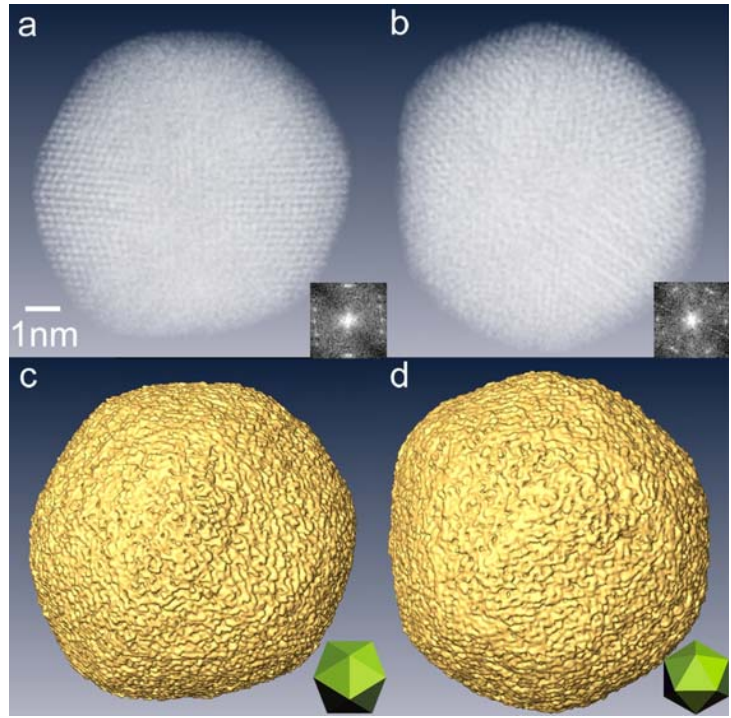

FIG. 3

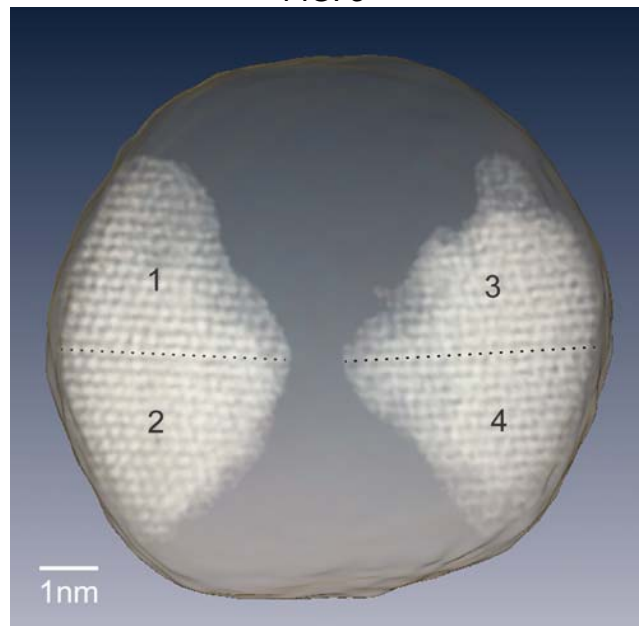

FIG. 4

# SUPPLEMENTARY INFORMATION
## Electron tomography at 2.4 Å resolution

M. C. Scott[1*], Chien-Chun Chen[1*], Matthew Mecklenburg[1*], Chun Zhu[1], Rui Xu[1], Peter Ercius[2], Ulrich Dahmen[2], B. C. Regan[1] & Jianwei Miao[1]

[1]*Department of Physics and Astronomy and California NanoSystems Institute, University of California, Los Angeles, CA 90095, USA.* [2]*National Center for Electron Microscopy, Lawrence Berkeley National Laboratory, Berkeley, CA 94720, USA.*
*These authors contributed equally to this work.

## Supplementary Methods

**Mathematical implementation of the EST method.** Before the iterative algorithm begins, EST requires that the projectional data be mapped onto the pseudopolar grid in reciprocal space. The EST method therefore begins with padding each projection with zeros (*i.e.,* embedding the experimental projection into a larger array of zeros) and calculating its oversampled Fourier slice on a pseudopolar grid (blue planes in Supplementary Fig. 2) using the fractional Fourier transform (FrFT)[33]. The FrFt varies the output sampling distance of the Fourier transform and is defined in the 1D case by

$$F_\alpha(k) = \sum_{x=-N}^{N-1} f(x) \exp(-\frac{i\pi\alpha kx}{N}) \ . \qquad (1)$$

Eq. (1) is equivalent to the standard 1D FFT but with an extra factor of $\alpha$ in the exponent. By choosing an appropriate value for $\alpha$, the projection data can be mapped on to the grid points of any line on the pseudopolar grid. The oversampling concept (*i.e.* sampling the Fourier slice at a frequency finer than the Nyquist interval)[34] has been widely used to solve the phase problem in coherent diffraction imaging[20,35-37]. In the EST method, oversampling does not provide extra information about the object, but allows the use of iterative algorithms to extract the correlated information within the projections. In the first iteration, the grid points outside the resolution circle (dashed line in Supplementary Fig. 1 left) and on the missing projections are set to zero. We also note that the reconstruction may be improved by supplying each missing projection with the average of its two neighboring projections as an initial input. Once this preprocessing step has occurred, the algorithm iterates back and forth between real and Fourier space, shown in Supplementary Fig. 2. The j[th] iteration consists of the following 5 steps:

i)   Apply the inverse PPFFT to the Fourier-space slices $F_j(\vec{k})$ , and obtain a real-space image, $f_j(\vec{r})$ . Our recent work has shown that the inverse PPFFT can be replaced by the adjoint PPFFT, allowing for faster convergence without compromising the accuracy [19].

ii)   A support (*S*) is determined based on the oversampling of the projections[34]. Outside the support, $f_j(\vec{r})$ is set to zero and inside the support, the negative values of $f_j(\vec{r})$ are set to zero. A new image, $f_j^{'}(\vec{r})$, is obtained ,

$$f_j^{'}(\vec{r}) = \begin{cases} 0 & \text{if } \vec{r} \notin S \text{ or } f_j(\vec{r}) < 0 \\ f_j(\vec{r}) & \text{if } \vec{r} \in S \text{ and } f_j(\vec{r}) \geq 0 \end{cases} \qquad (2)$$

iii)   Apply the PPFFT to $f_j^{'}(\vec{r})$ and obtain new Fourier slices, $F_j^{'}(\vec{k})$.

iv)   Calculate the Fourier slices for the (j+1)$^{\text{th}}$ iteration,

$$F_{j+1}(\vec{k}) = \begin{cases} F_j^{'\phi}(\vec{k}) & \text{for the missing projection angles } (\phi) \\ F_m^{\theta}(\vec{k}) & \text{for the measured projection angles } (\theta) \end{cases} \qquad (3)$$

where $F_j^{'\phi}(\vec{k})$ and $F_m^{\theta}(\vec{k})$ represent the missing and measured Fourier slices, and $\phi \cup \theta$ forms a complete set of angles for the pseudopolar grid.

v)   An $R_{recip}$ is calculated,

$$R_{recip} = \frac{\sum \left\| F_m^{\theta}(\vec{k}) \right| - \left| F_j^{'\theta}(\vec{k}) \right\|}{\sum \left| F_m^{\theta}(\vec{k}) \right|} \;, \qquad (4)$$

where $F_m^{\theta}(\vec{k})$ and $F_j^{'\theta}(\vec{k})$ represent the measured and j$^{\text{th}}$ calculated Fourier slices.

In our reconstructions, the algorithm is terminated after reaching a maximum number of iterations. To quantify the method, we project back the final 3D reconstruction to calculate a series of projections, which are quantified by an $R_{real}$,

$$R_{real} = \frac{\sum \left\| f_c^i(x,y) \right| - \left| f_m^i(x,y) \right\|}{\sum \left| f_m^i(x,y) \right|}, \qquad (5)$$

where $f_c^i(x,y)$ and $f_m^i(x,y)$ represent the calculated and measured projections in real space at tilt angle $i$.

**Supplementary Figures**

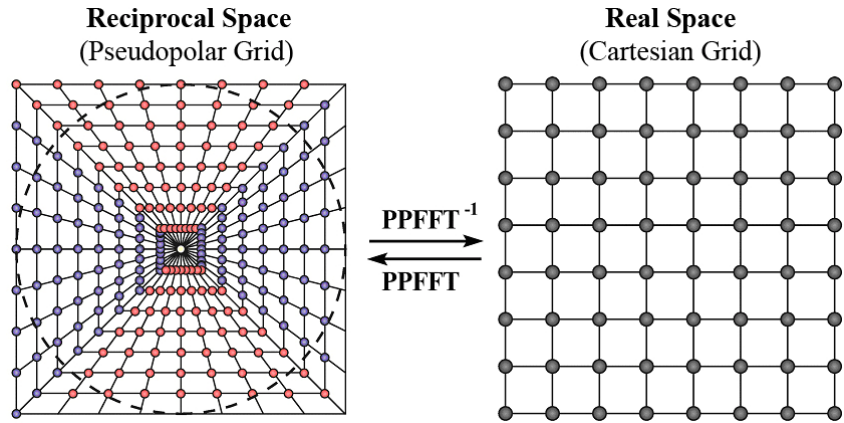

**Supplementary Fig. 1** Graphical relationship between the pseudopolar and Cartesian grids. For an $N \times N$ Cartesian grid, the corresponding pseudopolar grid is defined by a set of $2N$ lines, each line consisting of $2N$ grid points mapped out on $N$ concentric squares (left) with $N = 8$ in this example. The $2N$ lines are subdivided into a horizontal group (in blue) defined by $y = sx$, where $|s| \leq 1$, and a vertical group (in red) defined by $x = sy$, where $|s| \leq 1$. The horizontal and vertical groups are symmetric under the interchange of $x$ and $y$, and $\Delta s = 2/N$ [31]. The dashed circle on the pseudopolar grid represents the resolution circle. The grid points outside of the resolution circle cannot be obtained by applying the Fourier transform of the experimental projections[25].

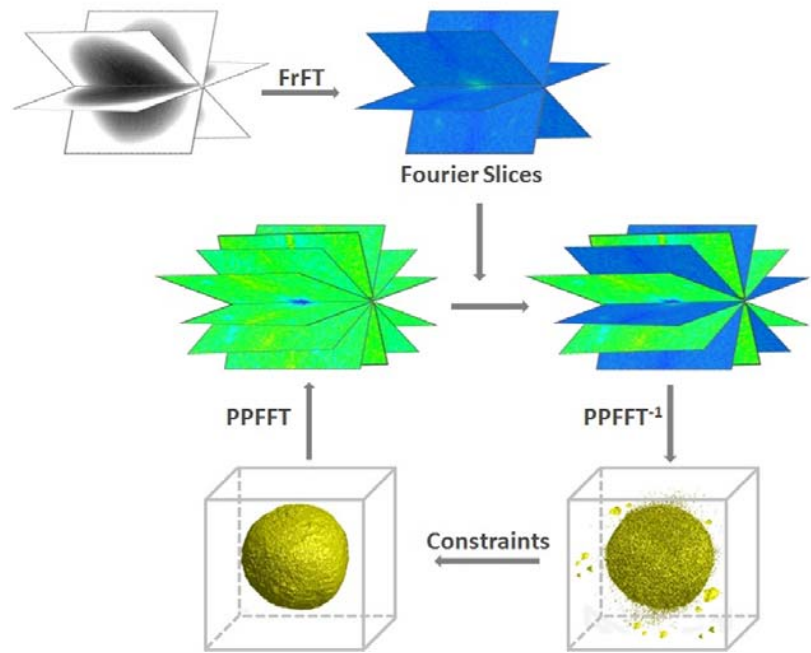

**Supplementary Fig. 2** Schematic layout of the iterative EST method. The measured projections are first converted to Fourier slices by the fractional Fourier transform (FrFT)[33]. The algorithm iterates back and forth between real and reciprocal space using the PPFFT and its inversion (Supplementary Fig. 1). In real space, the negative-valued voxels inside the support and the voxel values outside the support are set to zero (*i.e.* constraints are applied). In reciprocal space, the corresponding calculated slices are updated with the measured ones (in blue) and the remaining slices (in green) are unchanged. The algorithm is terminated after reaching a maximum number of iterations[25,16,18-20].

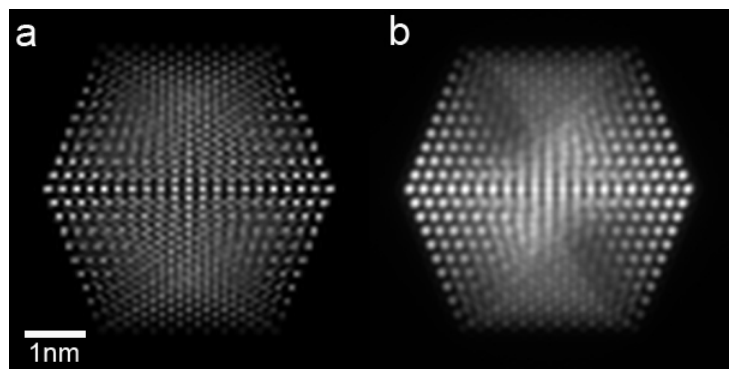

**Supplementary Fig. 3** Multislice calculations of a ~5 nm simulated Au nanoparticle with ideal icosahedral symmetry and a total of 3871 atoms. **a,** Projected Coulomb potential at 0°. **b,** 0° projection calculated by multislice STEM simulations (energy: 300 keV, spherical aberration: 1.2 mm, illumination semi-angle: 7.98 mrad, defocus: 48.6 nm, detector inner and outer angles: 13 and 78 mrad, pixel size: 0.37 Å). The particle was rotated by 1° each around the horizontal (X) and tilt (Y) axes to avoid the zone axis orientations and reduce the non-linear effects. The resolution in (**b**) was limited by the probe size (~1.5 Å), and the apparent increase of the atom size in the multislice projection was caused by diffraction and dynamical scattering effects in the nanoparticle.

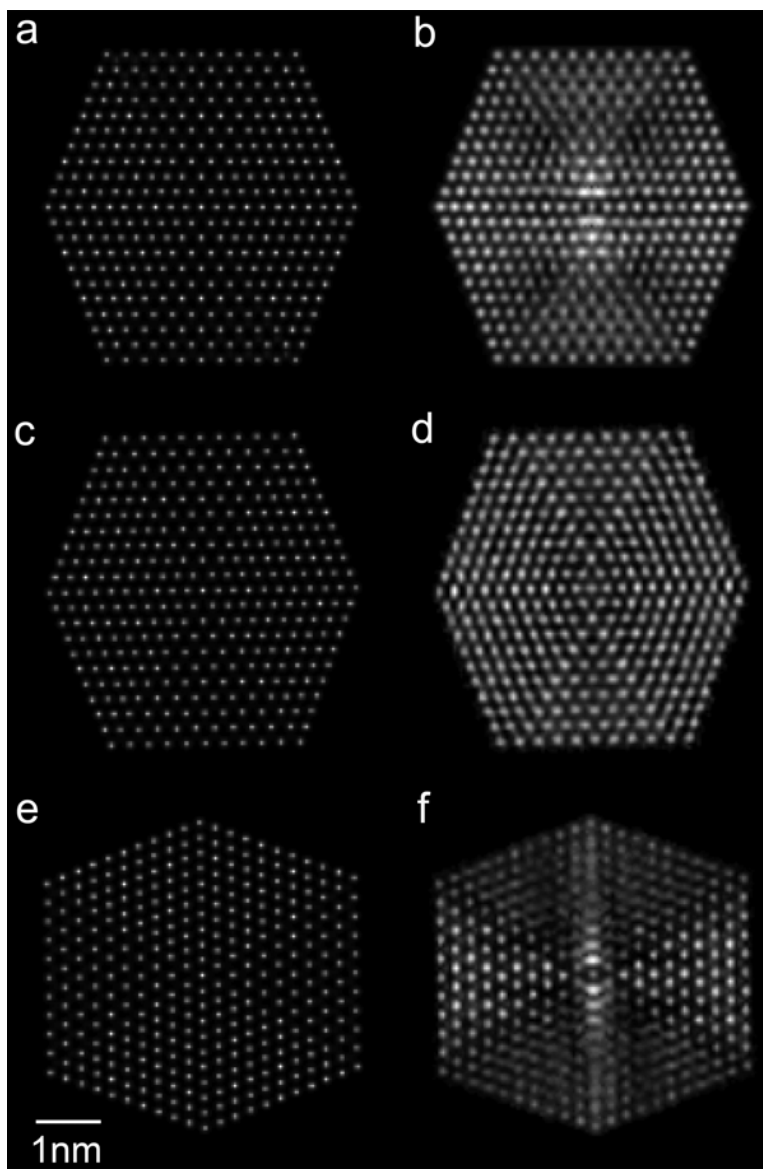

**Supplementary Figure 4.** EST reconstructions of the simulated Au nanoparticle (~5 nm) from a tilt series, calculated by multislice STEM simulations (energy: 300 keV, spherical aberration: 1.2 mm, illumination semi-angle: 7.98 mrad, defocus: 48.6 nm, detector inner and outer angles: 13 and 78 mrad, pixel size: 0.5 Å). To avoid the zone axis orientations and reduce the non-linear effects, the nanoparticle was rotated by 1° each around the horizontal (X) and tilt (Y) axes. The tilt series consists of 55 projections with a tilt range of ±72.6° and equal slope increments. To simulate experimental conditions, the tilt angles were continuously shifted from 0° to 0.5° over the process of the tilt series and the magnification of the images was continuously changed from 0 to 0.2%. The total dose of the tilt series is $6.1 \times 10^6$ e/Å$^2$ and Poisson noise was added to each projection. **a, c** and **e,** Three 2.5 Å thick central slices of the Coulomb potential of the simulated nanoparticle in the XY, XZ and YZ planes, where the Z-axis is the beam direction. **b, d** and **f,** The corresponding 2.5 Å thick slices in the XY, XZ and YZ planes reconstructed from 55 multislice STEM projections. Although the missing wedge

problem is not completely solved (the top and bottom parts in (**f**)) and the size of the reconstructed atoms is increased due to the non-linear and diffraction effects, the overall atomic positions and grain boundaries in the 3D reconstruction are consistent with the model.

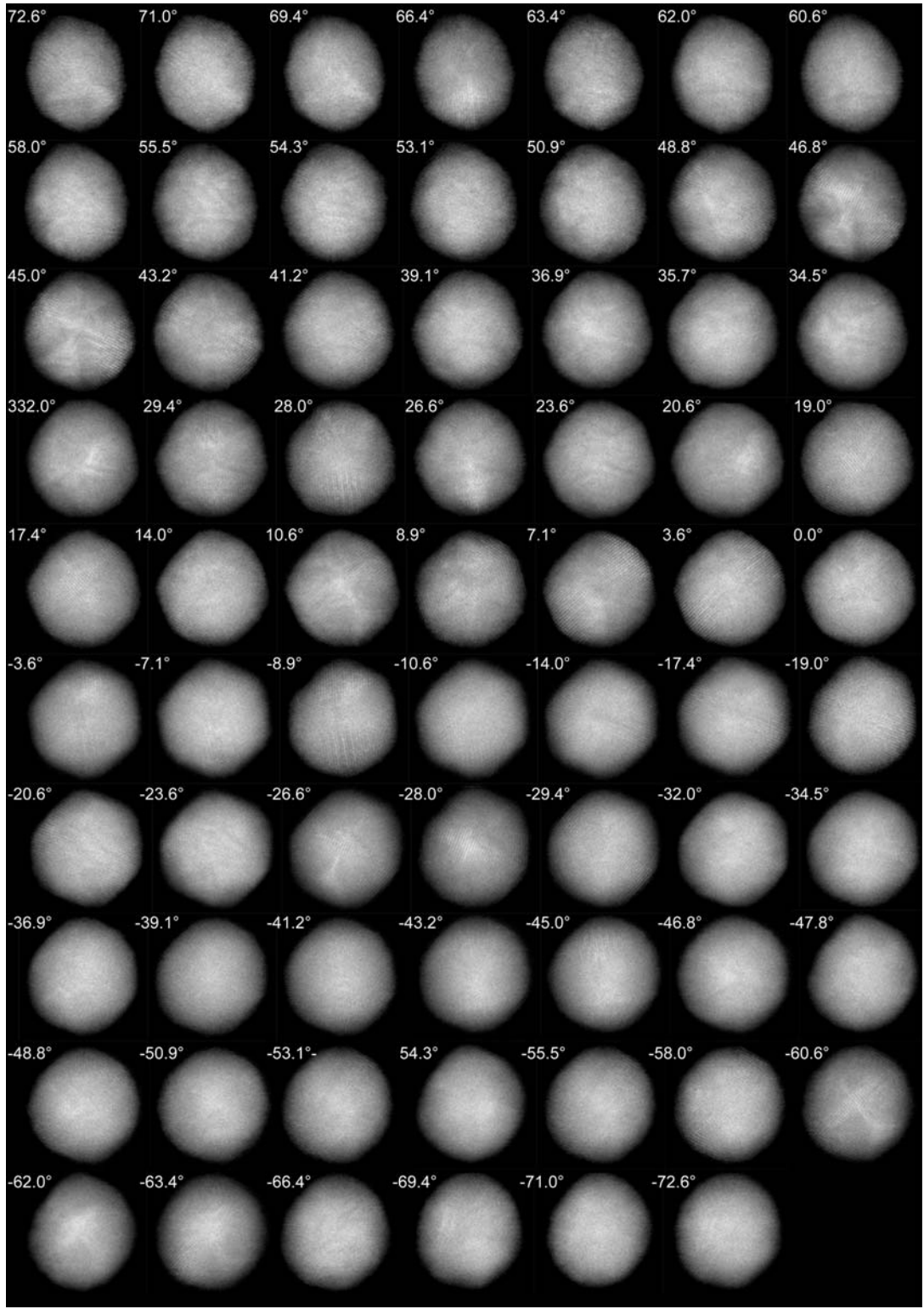

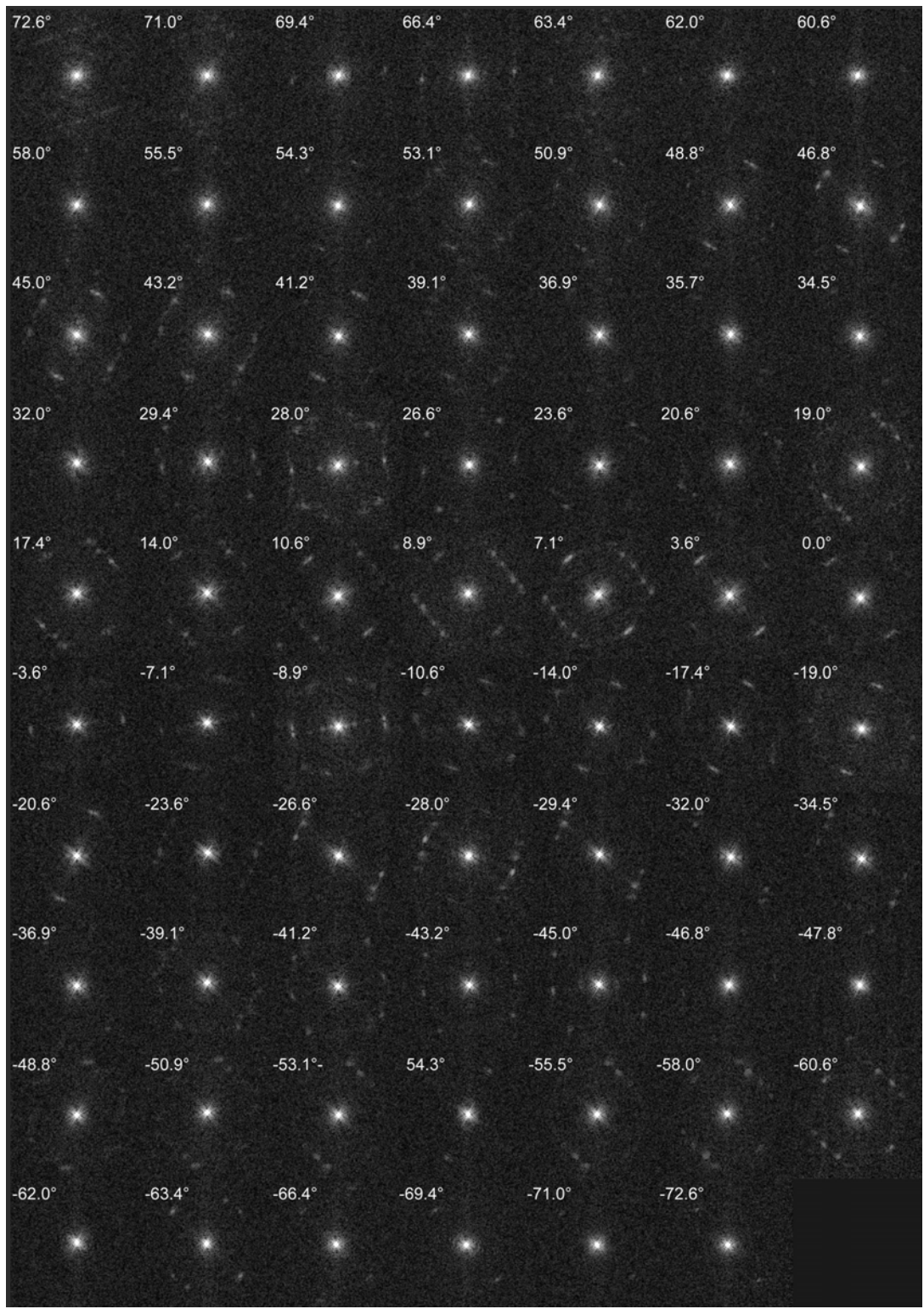

**Supplementary Figure 5.** Experimental tilt series of 69 projections and their Fourier transforms, acquired from a ~10 nm gold nanoparticle with the tilt axis along the vertical axis. Crystal lattices of the nanoparticle are visible in at least 58 projections.

The projections were acquired on an FEI Titan 80-300. The 300 keV electron beam, at spot 8 with a 50 μm C2 aperture, was focused to a probe with a probe current of ~70 pA, and rastered over the nanoparticle with a dwell time of 45 μs per pixel. The scattered electrons were captured by a Fischione Model 3000 ADF detector with angles between 13 and 78 mrad from the optical axis. The electron dose of this tilt series was estimated to be ~7.6×10$^6$ e/Å$^2$. Among all the 69 projections, the one at 7.1° is closest to a zone-axis orientation (about 17 mrad away from the 2-fold zone axis).

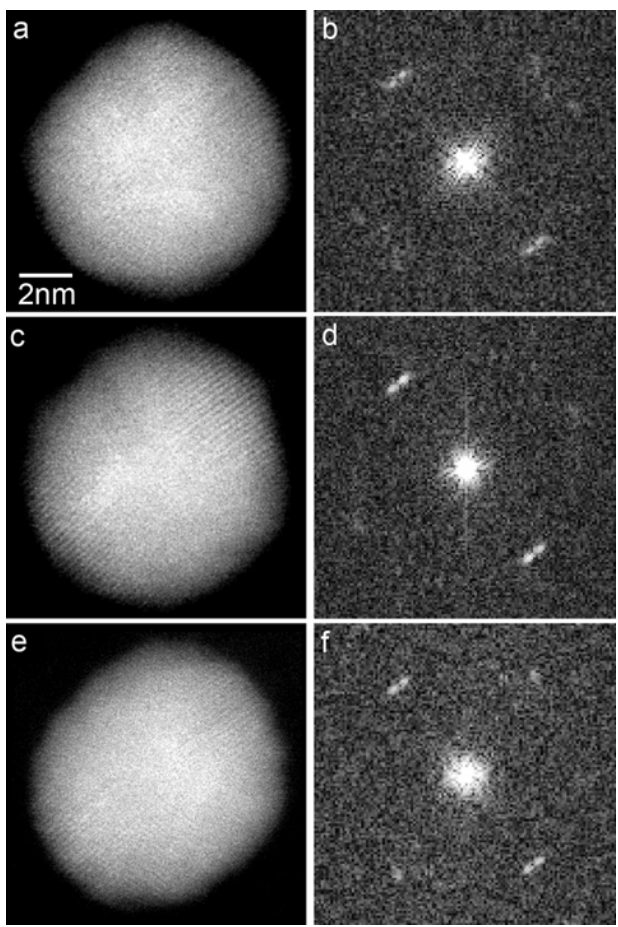

**Supplementary Figure 6.** Three 0° projections (**a,c,e**) and their Fourier transforms (**b,d,f**) measured during the acquisition of the tilt series (Supplementary Fig. 5) to monitor the effects of radiation damage. Although minor shape changes occurred, the overall crystal lattice structure of the gold nanoparticle remained consistent throughout the experiment. The minor shape change may contribute to a small degree of uncertainty in the overall shape of the reconstructed nanoparticle.

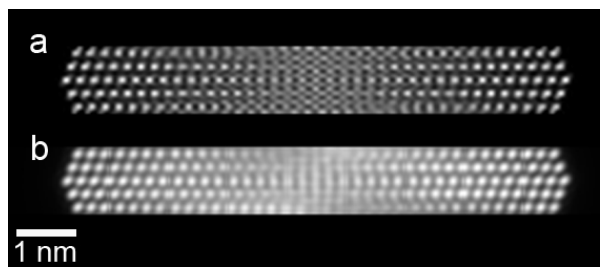

**Supplementary Fig. 7.** Multislice calculations for an 11.5 Å thick slab through the center of a ~10 nm simulated Au nanoparticle with ideal icosahedral symmetry and a total of 21127 atoms. **a,** Projected Coulomb potential at 0°. **b,** 0° projection calculated by multislice STEM simulations (energy: 300 keV, spherical aberration: 1.2 mm, illumination semi-angle: 7.98 mrad, defocus: 48.6 nm, detector inner and outer angles: 13 and 78 mrad, pixel size: 0.37 Å). The particle was rotated by 1° each around the horizontal (X) and tilt (Y) axes to avoid the zone axis orientations and reduce the non-linear effects. The resolution in (**b**) was limited by the probe size (~1.5 Å), and the apparent increase of the atom size in the multislice projection was caused by diffraction and dynamical scattering effects in the nanoparticle. As a proof of principle, we simulated only a 11.5 Å thick slab because calculating a full multislice STEM projection for a ~10 nm gold particle would take enormous computational power.

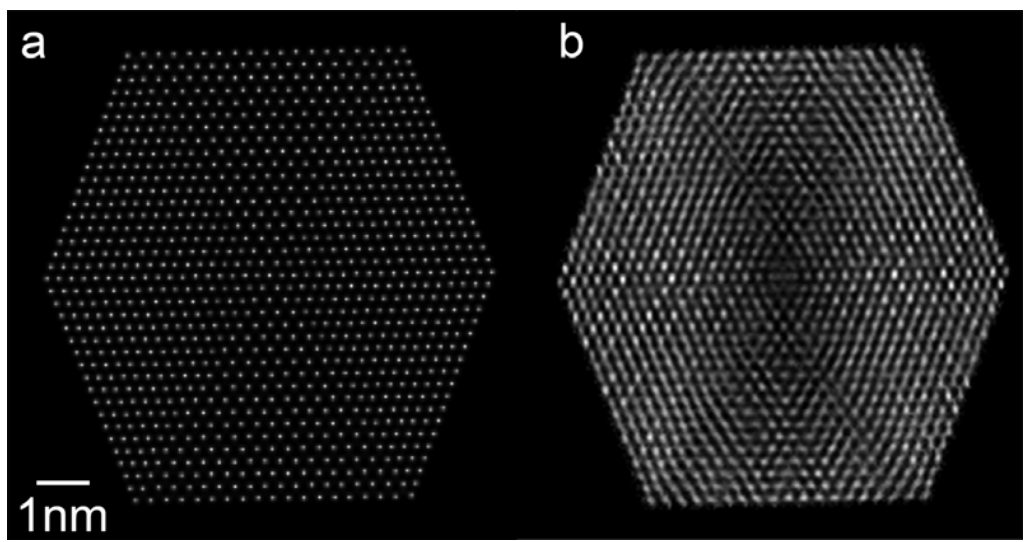

**Supplementary Fig. 8.** EST reconstruction of the ~10 nm simulated Au nanoparticle from a tilt series, calculated by multislice STEM simulations (energy: 300 keV, spherical aberration: 1.2 mm, illumination semi-angle: 7.98 mrad, defocus: 48.6 nm, detector inner and outer angles: 13 and 78 mrad, pixel size: 0.5 Å). The tilt series consists of 69 projections with a tilt range of ±72.6° and equal slope increments. To avoid the zone axis orientations and reduce the non-linear effects, the nanoparticle was rotated by 1° each around the horizontal (X) and tilt (Y) axes. To simulate experimental conditions, the tilt angles were continuously shifted from 0° to 0.5° over the process of the tilt series and the magnification of the images was continuously changed from 0 to 0.2%. The total dose of the tilt series is $7.6\times10^6$ e/$\text{Å}^2$ and Poisson noise was added to each projection. **a,** 2.5 Å thick central slice of the Coulomb potential in the XZ plane, where the Z-axis is the beam direction. **b,** 2.5 Å thick slice in the XZ plane reconstructed from 69 multislice STEM projections. The atomic positions and the internal grain boundaries are resolved except in very few places (including the origin) due to the non-linear effects in the projections. As a proof of principle, we only used a 2.5 Å thick slice to illustrate the EST reconstruction. Calculating a full tilt series for the ~10 nm gold nanoparticle by multislice STEM simulations would take enormous computational power.

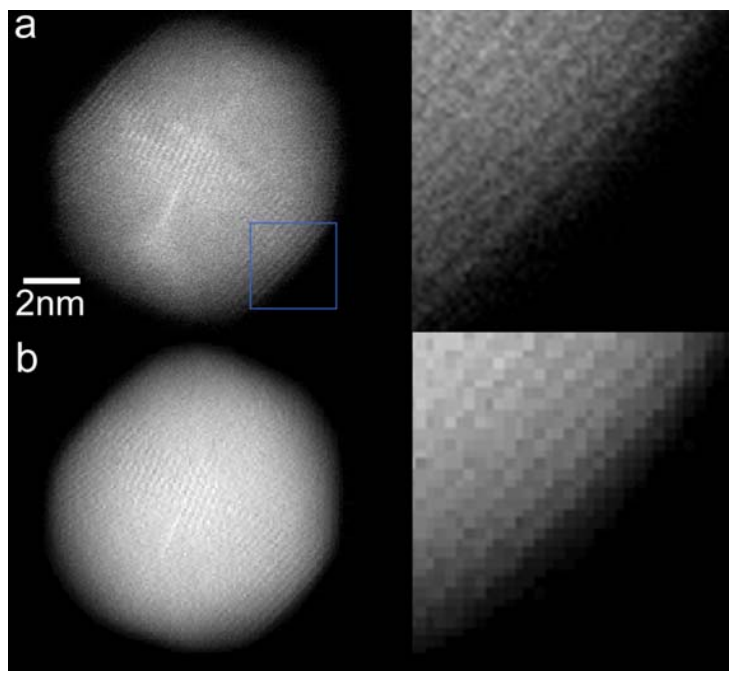

**Supplementary Figure 9.** Measured (**a**) and calculated (**b**) projections at -26.6° for the ~10 nm gold nanoparticle. The calculated projection was re-projected from the 3D reconstruction of 69 projections. The zoomed images indicate that, while there are some very minor differences between the two projections, the overall shape and lattice structure agree well.

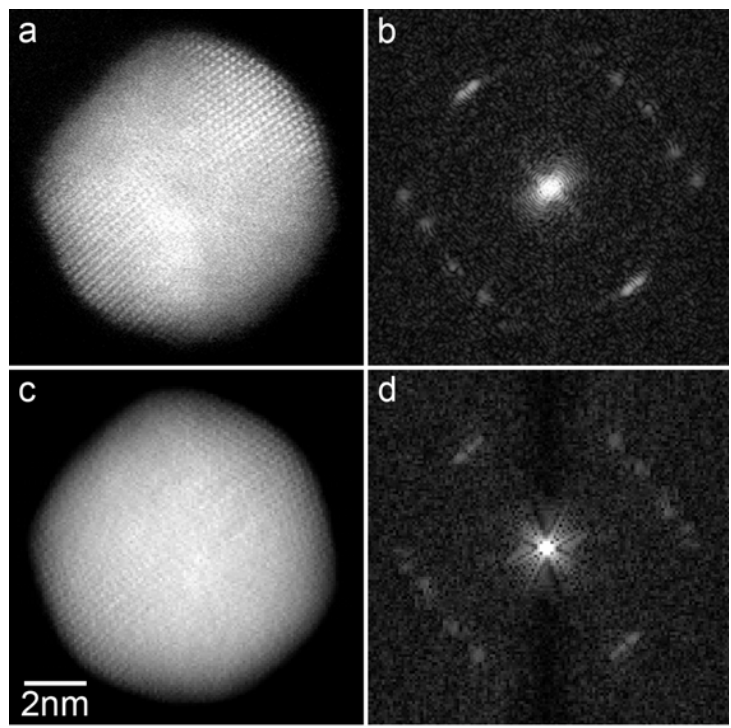

**Supplementary Figure 10.** Measured (**a**) and calculated (**c**) projections and their Fourier transforms (**b,d**) at 7.1°, where the calculated projection (**b**) was re-projected from a 3D reconstruction without using the measured projection (**a**). While the contrast of the lattice fringes and the Bragg peak intensity are different between (**a**), (**b**) and (**c**), (**d**), the overall shape and the lattice structure are in good agreement. In the reconstruction, the average of two neighboring projections at 3.6° and 8.9° was input as an initial guess for the projection at 7.1°, but was not used as a constraint in each iteration.

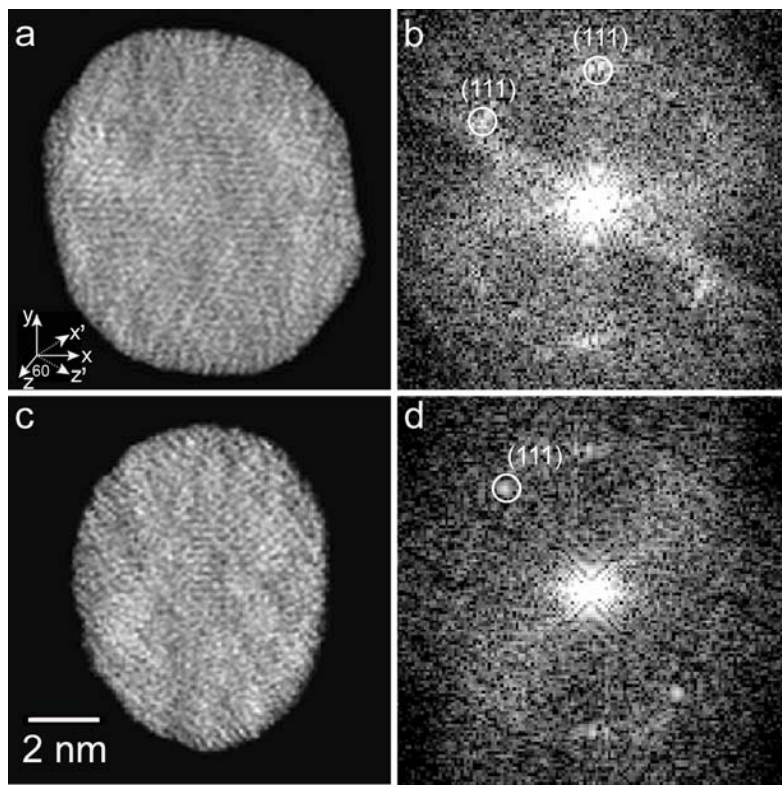

**Supplementary Figure 11. a** and **b**, A 3.36 Å slice in the X′Y plane and its Fourier transform, obtained from the experimentally reconstructed Au nanoparticle. **c** and **d**, A 3.36 Å slice in the Z′Y plane and its Fourier transform. The inset shows the direction of the X, Y, Z and X′, and Z′ axes, and the angle between planes ZY and Z′Y is ~60°, whereas the slices shown in Fig. 2a and c are in the XY and ZY planes, respectively. The crystal lattice structure is visible in the top and bottom areas in (**a**) and the top-right area in (**c**), but is not present in Figs. 2a and c.

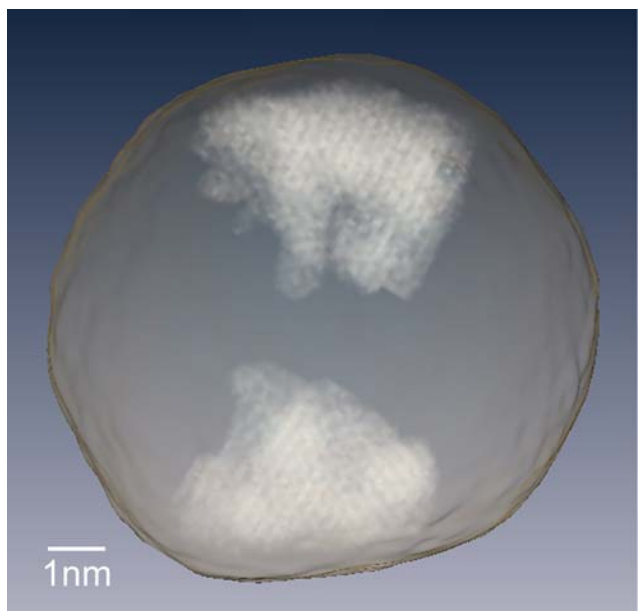

**Supplementary Figure 12.** 3D grains were identified in the top and bottom parts of the particle that is oriented at the 3-fold symmetry direction, whereas the particle in Fig. 4 is in the 2-fold symmetry orientation.

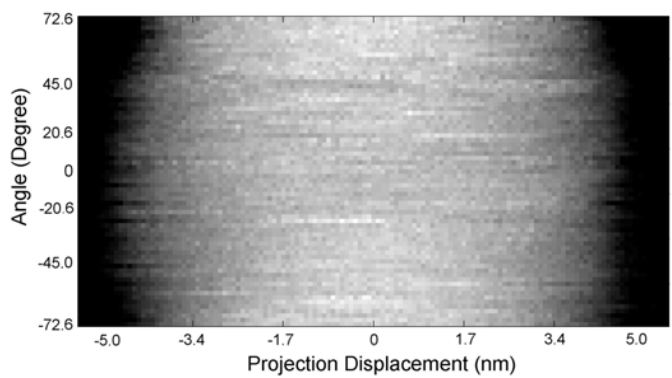

**Supplementary Figure 13.** A representative sinogram for the experimental tilt series of 69 projections acquired from the ~10 nm Au nanoparticle. The X-axis shows the distance to the tilt axis and the Y-axis shows the tilt angles for the projections. Unlike conventional tomography, the angular increments in EST are not constant. Thus the angles along the Y-axis are not equally distributed and smooth transitions between different angles are not expected. The sudden horizontal intensity jumps are mainly due to the lattice structure in the projections, and the rough edge is likely caused by the background and noise in the projections.

## Supplementary Tables

| Angles (°) | 72.6 | 71.0 | 69.4 | 66.4 | 63.4 | 62.0 | 60.6 |
|---|---|---|---|---|---|---|---|
| $R_{real}$ (%) | 7.1 | 6.2 | 6.4 | 10.8 | 10.2 | 5.9 | 6.2 |
| Angles (°) | 58.0 | 55.5 | 54.3 | 53.1 | 50.9 | 48.8 | 46.8 |
| $R_{real}$ (%) | 5.8 | 6.0 | 6.5 | 6.1 | 6.5 | 6.8 | 11.6 |
| Angles (°) | 45.0 | 43.2 | 41.2 | 39.1 | 36.9 | 35.7 | 34.5 |
| $R_{real}$ (%) | 14.0 | 11.4 | 8.0 | 5.8 | 6.1 | 6.9 | 6.4 |
| Angles (°) | 32.0 | 29.4 | 28.0 | 26.6 | 23.6 | 20.6 | 19.0 |
| $R_{real}$ (%) | 7.3 | 6.5 | 6.8 | 6.6 | 5.7 | 7.6 | 5.9 |
| Angles (°) | 17.4 | 14.0 | 10.6 | 8.9 | 7.1 | 3.6 | 0 |
| $R_{real}$ (%) | 5.6 | 5.0 | 7.4 | 7.7 | 10.5 | 5.7 | 5.3 |
| Angles (°) | -3.6 | -7.1 | -8.9 | -10.6 | -14.0 | -17.4 | -19.0 |
| $R_{real}$ (%) | 5.8 | 5.7 | 6.9 | 5.1 | 5.5 | 5.3 | 6.9 |
| Angles (°) | -20.6 | -23.6 | -26.6 | -28.0 | -29.4 | -32.0 | -34.5 |
| $R_{real}$ (%) | 5.4 | 6.1 | 6.9 | 8.6 | 5.6 | 5.0 | 4.8 |
| Angles (°) | -36.9 | -39.1 | -41.2 | -43.2 | -45.0 | -46.8 | -47.8 |
| $R_{real}$ (%) | 4.7 | 4.9 | 5.9 | 6.1 | 5.6 | 6.5 | 6.0 |
| Angles (°) | -48.8 | -50.9 | -53.1 | -54.3 | -55.5 | -58.0 | -60.6 |
| $R_{real}$ (%) | 7.4 | 6.7 | 5.9 | 6.5 | 6.0 | 6.6 | 7.2 |
| Angles (°) | -62.0 | -63.4 | -66.4 | -69.4 | -71.0 | -72.6 | Average |
| $R_{real}$ (%) | 6.9 | 7.8 | 5.6 | 6.6 | 5.5 | 6.6 | 6.7 |

**Supplementary Table 1.** To examine the reconstruction quality, we projected back the reconstructed 3D structure at the same experimental tilt angles to calculate 69 projections. An $R_{real}$ (Supplementary Methods) was calculated for each tilt angle. The average $R_{real}$ for all tilt angles is 6.7%.

## Supplementary Movies

**Supplementary Movie 1**. 3D volume rendering of the reconstructed gold nanoparticle.

**Supplementary Movie 2**. 3D iso-surface rendering of the reconstructed gold nanoparticle.

**Supplementary Movie 3**. 3D volume rendering of the four major grains determined from the reconstructed gold nanoparticle.